\documentclass[10pt, conference]{IEEEtran}
\IEEEoverridecommandlockouts

\usepackage{float}
\usepackage{cite}
\usepackage{amsmath,amssymb,amsfonts}
\usepackage{subcaption}
\usepackage{graphicx}
\usepackage{booktabs}
\usepackage{textcomp}
\usepackage{multirow}
\usepackage{xcolor}
\usepackage{xspace}
\usepackage{algorithm}
\usepackage{algpseudocode}

\def\BibTeX{{\rm B\kern-.05em{\sc i\kern-.025em b}\kern-.08em
    T\kern-.1667em\lower.7ex\hbox{E}\kern-.125emX}}

\begin{document}

\title{HybridFLow: SDN-Orchestrated Client Partitioning for Hybrid Federated Learning}

\author{
    \IEEEauthorblockN{
        Osama Abu Hamdan\IEEEauthorrefmark{1},
        Rabin Pandey\IEEEauthorrefmark{3},
        Hao Che\IEEEauthorrefmark{1},
        Engin Arslan\IEEEauthorrefmark{2},
        Md Arifuzzaman\IEEEauthorrefmark{3}
    }
    \IEEEauthorblockA{\IEEEauthorrefmark{1}%
        University of Texas at Arlington\\
        oma8085@mavs.uta.edu, hche@cse.uta.edu
    }
    \IEEEauthorblockA{\IEEEauthorrefmark{2}%
        Meta Platforms, Inc.\\
        enginarslan@meta.com
    }
    \IEEEauthorblockA{\IEEEauthorrefmark{3}%
        Missouri University of Science and Technology\\
        rp5dm@mst.edu, marifuzzaman@mst.edu
    }
}
% \author{}

\maketitle

\newcommand{\name}{\textit{HybridFLow}\xspace}
\newcommand{\smartflow}{\textit{SmartFLow}\xspace}
\newcommand{\fedasync}{\textit{FedAsync}\xspace}
\newcommand{\rfwd}{\textit{RFWD}\xspace}
\newcommand{\freecap}{\textit{FreeCap}\xspace}
\newcommand{\openflow}{\textit{OpenFlow}\xspace}

\begin{abstract}
Cross-silo Federated Learning (FL) enables geographically distributed institutions to collaboratively train machine learning models without sharing raw data. In wide-area deployments, however, communication delays often dominate round completion time and exacerbate the straggler effect. Hybrid FL addresses this challenge by combining synchronous and asynchronous client participation, but effective partitioning requires visibility into network conditions such as shared bottlenecks, link utilization, and path contention that individual clients cannot observe. We present \name, a closed-loop SDN-driven orchestration framework that integrates network-layer intelligence directly into hybrid FL. Leveraging the SDN controller's global topology view, \name generates calibrated per-client communication-time estimates before each training round and uses them to partition clients into synchronous and asynchronous groups while balancing round latency and update staleness. After each round, measured communication times are fed back to the controller to continuously refine future predictions. Experimental results across multiple network topologies show that \name reaches 80\% target accuracy 33--40\% faster than \smartflow and reduces average round duration by 30--40~seconds, while \fedasync fails to reach the target accuracy under non-IID data distributions.
\end{abstract}


\section{Introduction and Motivation}
\begin{figure}[t]
    \centering
    \includegraphics[width=0.7\linewidth]{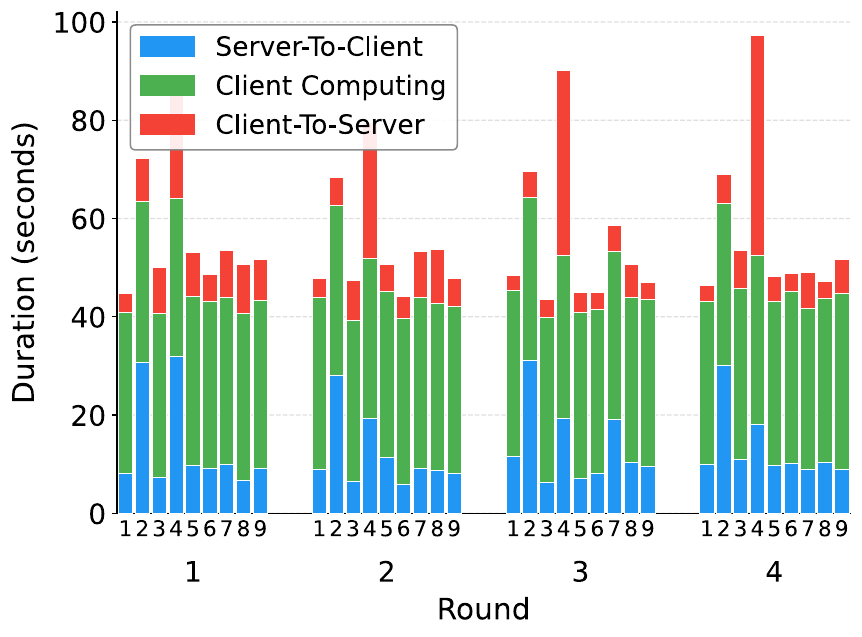}
    \caption{Straggler effect in synchronous FL. Round completion time is capped by the slowest client, forcing all faster participants to idle.}
    \label{fig:straggler}
    \vspace{-3mm}
\end{figure}

Cross-silo Federated Learning (FL) is an emerging class of distributed AI workload in which
geographically separated institutions collaboratively train a shared model without exchanging raw
data~\cite{intro2-mcmahan, intro4-bonawitz}. Unlike datacenter-local training, cross-silo FL
runs over wide-area networks with heterogeneous bandwidth, latency, and
congestion~\cite{intro1-kairouz, intro5-lyu}, making communication management the central
systems challenge. In synchronous FL, the server must wait for every participating client before
aggregating model updates, so training throughput is gated by the slowest network path. As
\figurename~\ref{fig:straggler} illustrates, this \emph{straggler effect} inflates per-round
completion time by over $47\%$ compared to optimized routing~\cite{smartflow}, with the penalty
growing with topology scale and link heterogeneity.

\begin{figure}[t]
    \centering
    \includegraphics[width=0.65\linewidth]{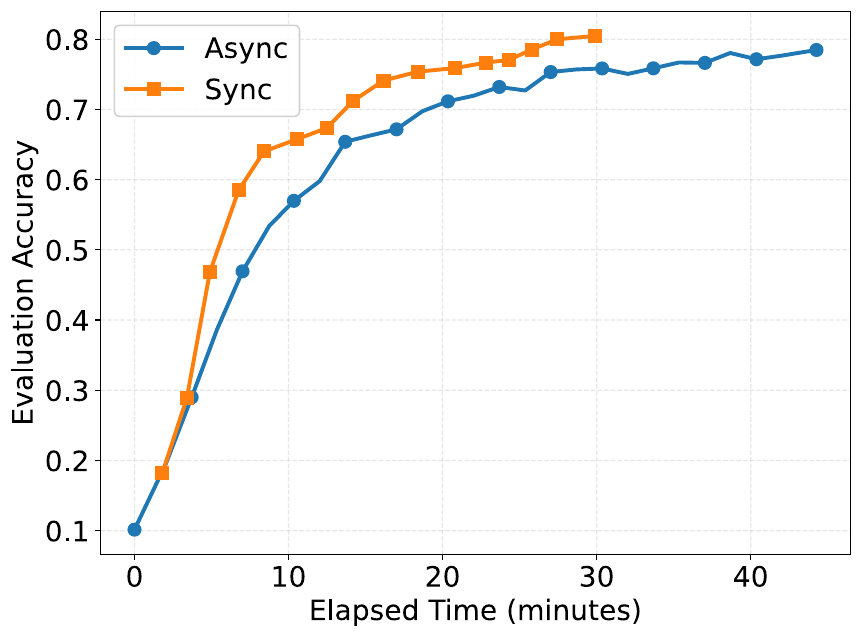}
    \caption{Convergence under non-IID data. Asynchronous FL eliminates the straggler bottleneck but fails to reach target accuracy under data heterogeneity, motivating a hybrid approach.}
    \label{fig:convergence}
    \vspace{-2mm}
\end{figure}

Fully asynchronous FL~\cite{fedasync} removes the synchronization barrier by integrating
updates as they arrive, but under non-IID data distributions fast clients dominate aggregation
while slow clients contribute stale gradients, degrading model convergence. As
\figurename~\ref{fig:convergence} shows, asynchronous FL fails to reach target accuracy under
data heterogeneity~\cite{asyncfeded, design-synchronous}. Hybrid FL addresses this trade-off
by assigning a synchronous subset of clients per round while allowing the remainder to contribute
asynchronously~\cite{design-synchronous}, preserving convergence quality without waiting on the
slowest paths.

Realizing the benefits of hybrid FL depends on correctly identifying which clients are likely
to hold up the synchronous phase before each round begins. In cross-silo settings, this is
fundamentally a network question: a client with capable hardware but a congested WAN path
behaves as a straggler regardless of its local compute speed, and that congestion is invisible
to the FL endpoints. End-host measurements expose only local symptoms; they cannot reveal
whether candidate paths share bottleneck links with other clients, or how background traffic
will evolve over the duration of a round. The partitioning decision therefore calls for a
signal that no individual client can generate on its own.

Software-Defined Networking provides exactly that signal. An SDN controller maintains a
topology-wide view of link utilization, latency, loss, and path contention across the entire
WAN~\cite{intro8-mendonca}. This global control-plane visibility is structurally distinct from
anything observable at the compute layer: it captures shared bottlenecks, cross-client path
interference, and network-wide congestion state that endpoint measurements cannot reach. Integrating this visibility into hybrid FL allows network-wide conditions to inform sync/async partitioning decisions.

\name is a closed-loop SDN orchestration framework for Hybrid Federated Learning in cross-silo deployments. It leverages real-time network telemetry from the SDN controller to generate calibrated per-client communication-time estimates and incorporates them into a cost-minimization algorithm that balances synchronous round latency against the staleness risk of asynchronous updates. By integrating topology-wide network intelligence directly into sync/async partitioning decisions, \name brings the network control plane into the Hybrid FL orchestration loop. The framework establishes a round-gated feedback mechanism in which communication-time estimates guide path assignment and client partitioning before each round, while measured transfer times continuously refine future estimates, closing the control loop without reactive mid-round rerouting.

This paper makes three contributions: (1) a closed-loop SDN-FL orchestration architecture that
translates network telemetry into per-round sync/async training-mode decisions; (2) a
network-aware hybrid partitioning algorithm that balances synchronous round latency against
asynchronous staleness risk using calibrated communication-time estimates from the SDN control
plane; and (3) an emulated cross-silo evaluation demonstrating reduced round time and faster
time-to-target accuracy under non-IID data distributions.

\section{Related Work}
Communication efficiency in FL has been extensively studied through gradient compression, quantization, and selective parameter updates~\cite{intro9-compression, intro8-quantization, intro10-dropping, konecny_communication_efficiency}. These approaches reduce communication volume and improve scalability, but operate primarily at the model and update levels. As a result, they do not directly address network-level factors such as path contention, routing inefficiencies, or time-varying congestion, which frequently dominate completion time in wide-area cross-silo FL deployments.

The synchronization bottleneck has motivated asynchronous and hybrid training paradigms. \fedasync~\cite{fedasync} demonstrated that a server can integrate updates without waiting for all clients, while AsyncFedED~\cite{asyncfeded} introduced distance-based weighting to reduce the impact of stale updates. Hybrid FL further balances convergence quality and training throughput by maintaining a synchronous subset of clients while allowing slower participants to contribute asynchronously~\cite{design-synchronous}. Existing hybrid approaches typically derive sync/async decisions from compute-layer information such as local training time, dataset size, gradient characteristics, or historical round latency. While these signals capture important aspects of client heterogeneity, they provide limited visibility into network-wide conditions such as shared bottlenecks, path contention, and evolving WAN congestion. In cross-silo environments, where communication time often dominates round completion time, incorporating network-layer information into the partitioning process represents a natural extension of existing hybrid FL strategies.

The intersection of SDN and FL has recently attracted increasing attention. Ma et al.~\cite{ma_survey} surveyed the challenges and opportunities of SDN-assisted FL systems, while Mahmod et al.~\cite{mahmod_freecap} demonstrated SDN-based bandwidth provisioning to mitigate link saturation during FL communication rounds. Mahmoud et al.~\cite{mahmoud_client_selection} further explored SDN-guided client selection as a mechanism for improving model quality. These studies highlight the value of network visibility and control for improving FL performance, but focus primarily on resource allocation and participant selection. Our prior work, \smartflow~\cite{smartflow}, leveraged ONOS-based telemetry and topology-aware routing to reduce synchronization time in cross-silo FL by up to $47\%$ relative to static routing. However, \smartflow retained a fully synchronous training model and therefore remained susceptible to network-induced stragglers.

\name extends prior SDN-assisted FL approaches by using SDN telemetry not only to optimize communication paths, but also to inform application-level training orchestration. Specifically, network-derived communication-time estimates are incorporated directly into the hybrid sync/async partitioning process, enabling the network control plane to participate in per-round training-mode decisions. To the best of our knowledge, prior SDN-assisted FL studies have not used topology-wide network telemetry as a control signal for hybrid synchronous/asynchronous client partitioning in cross-silo federated learning.

\section{System Overview}
\begin{figure*}[t]
    \centering
    \includegraphics[width=0.95\linewidth]{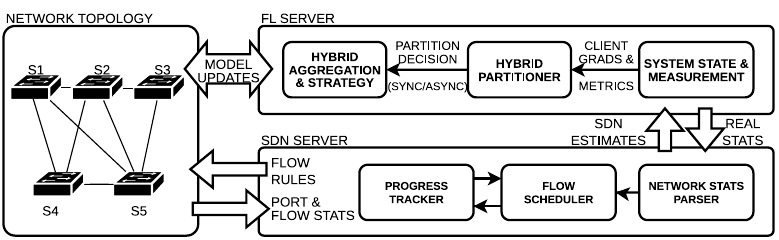}
    \caption{\name architecture. The SDN controller produces calibrated per-client
    communication time estimates via the \emph{Flow Scheduler} and \emph{Progress Tracker};
    the FL server consumes them to partition clients and orchestrate hybrid aggregation.
    One \emph{Estimate Request}--\emph{Measurement Report} exchange per round closes the
    feedback loop.}
    \label{fig:architecture}
    \vspace{-2mm}
\end{figure*}

\name is organized across two cooperating but independent layers, as illustrated in
\figurename~\ref{fig:architecture}: an \emph{SDN Controller} layer
implemented in Java on the ONOS platform~\cite{onos}, and a \emph{Federated Learning} layer
implemented in Python using the Flower framework~\cite{intro12-flower}. A lightweight ZeroMQ
messaging interface connects the two layers. The SDN layer handles network monitoring, path
scoring, and per-client communication time estimation; the FL layer handles client partitioning,
training orchestration, and model aggregation. Neither layer accesses the internal state of the
other; the messaging protocol serves as the sole interface.

The remainder of this section presents the system in three parts. First, the SDN layer
establishes the network-level infrastructure that produces per-client communication time
estimates. Second, the FL layer consumes those estimates to partition clients into synchronous
and asynchronous groups and to orchestrate hybrid training. Third, the data exchange protocol
that connects the two layers closes the feedback loop.

% -----------------------------------------------------------------------
\subsection{SDN Layer}

The SDN layer builds on the ONOS controller and retains the \emph{Stats Parser},
\emph{Client Store}, and \emph{Link Store} components introduced in
\smartflow~\cite{smartflow}. These components continuously collect \openflow statistics from
network devices, maintaining per-link estimates of available capacity $F_l^{\text{free}} =
B_l - U_l$, one-way latency $\lambda_l$, and packet loss probability $p_l$, all smoothed by
an Exponentially Weighted Moving Average (EWMA) over the three most recent
measurements~\cite{latency_and_loss}. The \emph{Flow Scheduler} and \emph{Progress Tracker}
from \smartflow are retained in \name and extended to support the hybrid FL workflow.

\subsubsection{The Path Assignment Problem}

Assigning network paths to FL clients is a non-trivial decision because: (i) multiple candidate
paths may connect each client to the server, with widely varying bandwidth, latency, and loss
characteristics; (ii) paths share links, so assigning one client to a high-quality path reduces
the available capacity for subsequent clients; and (iii) the quality of any assignment depends on
real-time link conditions that change between rounds. The \emph{Flow Scheduler} takes the
current network state as input and produces, for every client in both the server-to-client (S2C)
and client-to-server (C2S) directions, a single assigned path together with a calibrated
transfer time estimate.

To quantify path quality, the \emph{Flow Scheduler} evaluates each candidate path
$P = (l_1, \dots, l_m)$ (a sequence of directed links excluding edge links) on three
complementary metrics.

\textbf{Projected Fair Share.}
The estimated per-client bandwidth on link $l$, assuming the incoming client joins the
$n_l$ flows already traversing it, is:
\begin{equation}
\text{FairShare}_l^{\text{proj}} = \frac{F_l^{\text{free}}}{n_l + 1}.
\end{equation}
The path-level fair share equals the bottleneck link's share:
\begin{equation}
\text{FairShare}_P^{\text{proj}} = \min_{l \in P}\, \text{FairShare}_l^{\text{proj}}.
\end{equation}
$n_l$ is initialized from the per-link active-flow count reported by OpenFlow statistics and
incremented by the \emph{Flow Scheduler} each time a path using link $l$ is assigned during
the current scheduling sweep, so it reflects both existing background traffic load and the
cumulative capacity consumed by earlier assignments within the same round.

\textbf{Effective RTT.}
The round-trip time of path $P$ sums forward and reverse latencies across every link:
\begin{equation}
\text{RTT}_P = \sum_{l \in P} \bigl(\lambda_l + \lambda_{\bar{l}}\bigr),
\end{equation}
where $\bar{l}$ denotes the reverse direction of link $l$. This formulation accurately
models TCP behavior: data packets traverse the forward path while acknowledgment segments
return along $\bar{l}$, so the per-link contribution to round-trip time is precisely
$\lambda_l + \lambda_{\bar{l}}$~\cite{tcp_rfc5681}.

\textbf{End-to-End Packet Loss.}
Under the assumption of independent link failures:
\begin{equation}
p_P = 1 - \prod_{l \in P}(1 - p_l).
\end{equation}

\textbf{Effective Score.}
Combining these metrics into a Mathis-variant throughput estimate~\cite{adj_rtt}:
\begin{equation}
\text{EffScore}(P) =
  \frac{\text{FairShare}_P^{\text{proj}}}
       {\text{RTT}_P \cdot \sqrt{p_P + \varepsilon}},
\label{eq:effscore}
\end{equation}
where $\varepsilon$ is a small positive value that prevents division by zero when the
measured packet loss on a path is negligible. A higher score indicates a path that delivers
more throughput per unit time. Dividing the model size $D$ (in megabits) by
$\text{EffScore}(P)$ yields a raw transfer time estimate in seconds.

While the classical Mathis formulation models steady-state TCP behavior under random drop
conditions, we employ $\text{EffScore}(P)$ strictly as a comparative, network-layer utility
proxy to rank candidate paths prior to transmission. Any real-world deviations from
steady-state dynamics are captured and corrected downstream by the \emph{Progress Tracker}'s
EWMA feedback loop, which adjusts raw estimates toward observed per-client transfer times
across successive rounds.

The \emph{Flow Scheduler} interface is implementation-agnostic: any algorithm that maps the
current network state to per-client path assignments and time estimates can serve as the
scheduler. For instance, constraint programming formulations~\cite{smartflow} and reinforcement
learning-based solvers are both viable alternatives. In \name, the current implementation uses
a greedy allocation that, for each direction
$d \in \{\text{S2C},\, \text{C2S}\}$, sorts clients in ascending order of their best
available adjusted score so that the most constrained clients receive first choice of paths,
assigns each client to its highest-scoring path, and increments the active-flow count on every
link of the assigned path to reflect the capacity consumed by that assignment.

Paths are assigned once per round, at the start of each training round; no dynamic reassignment
occurs during the transfer phase. Because the \emph{Hybrid Partitioner} divides clients into
synchronous and asynchronous groups at the beginning of each round
(Section~\ref{sec:fl_layer}), fixing path assignments at the same point is a natural
simplification that keeps the scheduling calculation tractable while ensuring consistency
between routing and partitioning decisions.

\subsubsection{Correction Factors and Adjusted Scoring}

Raw effective scores are throughput-proxy values, not precise throughput predictions. To
compensate for systematic bias, the \emph{Progress Tracker} maintains a multiplicative
correction factor $\text{CF}(c, d, P)$ for each (client, direction, path) triple, updated via
EWMA after every round:
\begin{equation}
r_{c,d,P}^{(t)} = \frac{T_{c,d}^{\text{real},(t)}}{T_{c,d}^{\text{raw},(t)}},
\quad T_{c,d}^{\text{raw}} = \frac{D}{\text{EffScore}(P)},
\end{equation}
\begin{equation}
\text{CF}_{c,d,P}^{(t)} = \beta \cdot r^{(t)} + (1-\beta)\cdot\text{CF}^{(t-1)},
\label{eq:cf}
\end{equation}
where $\beta$ is the EWMA smoothing parameter that controls how rapidly the correction factor
adapts to new observations. When per-client history is unavailable, the system falls back to
aggregate correction data; if no historical data exists at all, the correction factor defaults
to $1.0$. The adjusted score and final time estimate are:
\begin{equation}
\text{AdjScore}(P, c, d) = \frac{\text{EffScore}(P)}{\text{CF}(c, d, P)},
\end{equation}
\begin{equation}
\hat{T}_{c,d} = \frac{D}{\text{AdjScore}(P^*_{c,d}, c, d)}.
\label{eq:time_est}
\end{equation}
Equation~\eqref{eq:time_est} yields a calibrated communication time estimate in seconds that
improves with each round as the correction factor converges.

% -----------------------------------------------------------------------
\subsection{FL Layer}
\label{sec:fl_layer}

The FL layer receives per-client communication time estimates from the SDN layer and uses them
to partition clients, orchestrate training, and aggregate model updates.

\subsubsection{Client State and Completion Time Estimation}

The client state record tracks per-client state across rounds, including communication
estimates and EWMA-smoothed computation times. The estimated total completion time for
client $c$ before round $t$ is:
\begin{equation}
T_c = \hat{T}_{c,\text{S2C}} + \hat{T}_{c,\text{C2S}} + \bar{t}_c^{\text{comp}},
\label{eq:Tc}
\end{equation}
where $\hat{T}_{c,\text{S2C}}$ and $\hat{T}_{c,\text{C2S}}$ are the SDN-provided
communication time estimates, and $\bar{t}_c^{\text{comp}}$ is the EWMA of client $c$'s
observed local computation times.

\subsubsection{Gradient Importance Tracking}

The gradient importance tracker maintains an EMA-smoothed estimate of each client's gradient
importance. After each round, it records the L2 norm of the parameter difference
$\|g_c^{\text{last}}\| = \|w_c - w_{\text{global}}\|_2$ for every client $c$ that contributed
an update. The global average is then updated:
\begin{equation}
\bar{g}_{\text{new}} = \delta\cdot\bar{g}_{\text{old}} + (1-\delta)\cdot
\frac{1}{|R|}\sum_{c\in R} \|g_c^{\text{last}}\|,
\end{equation}
where $\delta$ is a smoothing parameter that controls how quickly the global average adapts to
recent gradient magnitudes, and $R$ is the set of all clients that contributed updates in the
current round (both synchronous and asynchronous). The per-client importance weight is:
\begin{equation}
w_c^{\text{imp}} = \frac{\|g_c^{\text{last}}\|}{\bar{g}}.
\label{eq:imp}
\end{equation}
In the first round, $\|g_c^{\text{last}}\| = \bar{g} = 1.0$, so all clients begin with
uniform importance $w_c^{\text{imp}} = 1.0$. Clients with above-average gradient norms
($w_c^{\text{imp}} > 1$) contribute more to model improvement; assigning them to the
asynchronous group carries a higher staleness penalty, biasing the partitioner toward
keeping them synchronous.

\subsubsection{Network-Aware Client Partitioning}

The \emph{Hybrid Partitioner} solves the core optimization problem of \name: given estimated
completion times $\{T_c\}$ and importance weights $\{w_c^{\text{imp}}\}$, partition clients
into a synchronous set $\mathcal{S}$ and an asynchronous set
$\mathcal{A} = C\setminus\mathcal{S}$, minimizing a cost function that balances round latency
against staleness risk. The cost function combines a sync latency term and an async penalty
term, each normalized to $[0, 1]$.

The sync latency term measures how much the synchronous round duration exceeds its minimum
possible value:
\begin{equation}
J_{\text{sync}}(\mathcal{S}) =
  \frac{\max_{c\in\mathcal{S}} T_c - T_{\min}}{T_{\max} - T_{\min}},
\label{eq:jsync}
\end{equation}
where $T_{\min} = T_{c_{k_{\min}}}$ is the sync latency at the minimum sync set size
$k_{\min}$, and $T_{\max} = T_{c_n}$ is the slowest client's estimated time. This min-max
normalization ensures the term spans the full $[0, 1]$ range across all candidate split points.

The async penalty term captures the staleness risk of the asynchronous group. For each
asynchronous client $c$, gradient importance $w_c^{\text{imp}}$ is weighted by a quadratic
staleness proxy, penalizing clients whose completion time far exceeds the synchronous boundary:
\begin{equation}
J_{\text{async}}(\mathcal{S}) =
  \frac{\displaystyle\sum_{c\in\mathcal{A}} w_c^{\text{imp}} \cdot
    \left(\frac{T_c}{\max_{s\in\mathcal{S}} T_s}\right)^{\!2}}{P_{\max}},
\label{eq:jasync}
\end{equation}
where $P_{\max}$ is the raw penalty at the split with the largest async set, normalizing the
term to $[0, 1]$. The normalization constants $T_{\min}$, $T_{\max}$, and $P_{\max}$ are
computed once at the beginning of each sweep and remain fixed for its duration.

The total cost combines the two terms as a convex combination:
\begin{equation}
J(\mathcal{S}) =
  (1 - \lambda)\, J_{\text{sync}}(\mathcal{S})
  + \lambda\, J_{\text{async}}(\mathcal{S}),
\label{eq:J}
\end{equation}
subject to $|\mathcal{S}| \geq M_{\text{sync}}^{\min}$. The trade-off parameter
$\lambda \in [0, 1]$ governs the balance: $\lambda = 0$ minimizes sync latency alone (fastest
rounds), $\lambda = 1$ minimizes the async penalty alone (safest convergence), and the default
$\lambda = 0.5$ weights both objectives equally. Because both constituent terms are normalized,
$\lambda$ has scale-independent semantics.

Because all clients are sorted by $T_c$ and the optimal split is a contiguous prefix of this
ordering, the minimizer can be found by a greedy sweep in $O(n\log n)$ time, as formalized in
Algorithm~\ref{alg:partitioner}.

\begin{algorithm}[t]
\caption{\emph{Hybrid Partitioner}: Greedy Sweep}
\label{alg:partitioner}
\begin{algorithmic}[1]
\Require Clients $C$, times $\{T_c\}$, weights $\{w_c^{\text{imp}}\}$,
         $\lambda$, $k_{\min}$
\State Sort $C$ by $T_c$: $T_{c_1} \leq \dots \leq T_{c_n}$
\State $T_{\min} \gets T_{c_{k_{\min}}}$;\;
       $T_{\max} \gets T_{c_n}$;\;
       $P_{\max} \gets \text{AsyncPenalty}(k_{\min})$
\State $J^* \gets \infty$;\; $k^* \gets k_{\min}$
\For{$k = k_{\min}$ \textbf{to} $n$}
    \State $J_s \gets (T_{c_k} - T_{\min}) \,/\, (T_{\max} - T_{\min})$
    \State $J_a \gets \text{AsyncPenalty}(k) \,/\, P_{\max}$
    \State $J \gets (1-\lambda) \cdot J_s + \lambda \cdot J_a$
    \If{$J < J^*$}
        $J^* \gets J$;\; $k^* \gets k$
    \EndIf
\EndFor
\State \Return $\mathcal{S}^* = \{c_1,\dots,c_{k^*}\}$,\;
               $\mathcal{A}^* = \{c_{k^*+1},\dots,c_n\}$
\end{algorithmic}
\end{algorithm}

\subsubsection{Hybrid Aggregation}

Once the \emph{Hybrid Partitioner} determines $\mathcal{S}$ and $\mathcal{A}$, the hybrid
server dispatches all asynchronous clients to background training immediately, then blocks on the
completion of all synchronous clients. Completed asynchronous updates arrive at varying times
and are deposited into a thread-safe buffer. Aggregation then proceeds in two stages.

\textbf{Stage~1: Synchronous Aggregation.}
The server aggregates synchronous results using the configured aggregation strategy. \name
is compatible with any aggregation algorithm (e.g., FedAvg, FedProx); the choice of strategy
is orthogonal to the partitioning and scheduling mechanisms described here.

\textbf{Stage~2: Asynchronous Integration.}
\name integrates buffered asynchronous updates following the staleness-aware aggregation
framework introduced by \fedasync~\cite{fedasync}. Each buffered update carries a staleness
measure $\tau_c = t_{\text{current}} - t_{\text{submission},c}$ (in rounds). The per-client
effective mixing coefficient combines a configurable base mixing rate with a staleness decay:
\begin{equation}
\alpha_{\text{eff},c} =
  \alpha_{\text{base}}
  \cdot
  f_{\text{staleness}}(\tau_c),
\label{eq:alpha_eff}
\end{equation}
where $\alpha_{\text{base}} \in [0, 1]$ controls the maximum influence any single asynchronous
update can exert. The staleness decay function $f_{\text{staleness}}$ can be configured as
\emph{polynomial} $(\tau+1)^{-a}$, \emph{exponential} $e^{-a\tau}$, or
\emph{constant} $1.0$; the default is polynomial with $a = 0.5$,
matching the \fedasync formulation~\cite{fedasync}.

The buffered updates are first combined into a staleness-weighted average:
\begin{equation}
w_{\text{avg}} =
  \frac{\displaystyle\sum_{c \in \mathcal{A}} \alpha_{\text{eff},c}\, w_c}
       {\displaystyle\sum_{c \in \mathcal{A}} \alpha_{\text{eff},c}}.
\label{eq:wavg_async}
\end{equation}

This average is then merged with the synchronous model $w_{\text{sync}}$ through a collective
coefficient that separates data fraction from staleness quality:
\begin{align}
\alpha_{\text{coll}} &=
  \frac{N_{\mathcal{A}}}{N_{\mathcal{A}} + N_{\mathcal{S}}}
  \;\times\;
  \frac{\overline{\alpha_{\text{eff}}}}{\alpha_{\text{base}}},
\label{eq:alpha_coll}\\
w_{\text{new}} &=
  (1 - \alpha_{\text{coll}})\cdot w_{\text{sync}}
  + \alpha_{\text{coll}}\cdot w_{\text{avg}}.
\label{eq:async_merge}
\end{align}
The first factor reflects the asynchronous cohort's share of the total training data; the second
factor ($\in [0,1]$) scales that share proportionally to the mean staleness quality of the
batch. Their product always lies in $[0, 1]$, requiring no clamping heuristics.

\subsubsection{Async Dispatch and Re-Dispatch}

Each asynchronous client is submitted to a thread-pool executor with a per-client queue depth of
one: if a client already has an in-flight task, the dispatcher skips it. When an asynchronous
client completes, the system deposits its update into the buffer and makes a re-dispatch
decision based on round timing. If the client finished within the same round it was dispatched,
it is \emph{not} re-dispatched, because it may belong in the synchronous group in the next
round's partitioning. If the client finished after the server has already advanced to a new
round, the system re-dispatches it with the latest global model to keep it productive until the
next partitioning decision.

% -----------------------------------------------------------------------
\subsection{FL-SDN Data Exchange Protocol}
\label{sec:zmq}

\name coordinates its two layers through a proactive, round-gated data exchange protocol. The
protocol comprises two message types exchanged once per round via ZeroMQ, ensuring that all
path assignments and time estimates are finalized before any data transfer occurs.

\textbf{Estimate Request.}
At the start of round $t$, the FL server issues an estimate request to the SDN controller
containing the current round number and the list of available client identifiers. The SDN
controller invokes the \emph{Flow Scheduler} to run path assignment and returns a map
$\{c_i \mapsto [\hat{T}_{c_i,\text{S2C}},\; \hat{T}_{c_i,\text{C2S}}]\}$ of adjusted
communication time estimates for all clients. The FL server applies these estimates to
compute per-client completion times via Equation~\eqref{eq:Tc}.

\textbf{Measurement Report.}
After the round concludes, the FL server sends a map
$\{c_i \mapsto [T_{c_i,\text{S2C}}^{\text{real}},\; T_{c_i,\text{C2S}}^{\text{real}}]\}$
of measured communication times to the SDN controller. The \emph{Progress Tracker} uses these
measurements to update the EWMA correction factors per Equation~\eqref{eq:cf}, closing the
feedback loop and improving estimation accuracy in subsequent rounds.

% -----------------------------------------------------------------------
\subsection{End-to-End Round Execution}

Algorithm~\ref{alg:hybridflow} summarizes the complete per-round execution of \name,
integrating the SDN and FL layers through the data exchange protocol described in
Section~\ref{sec:zmq}.

\begin{algorithm}[t]
\caption{\name: Per-Round Execution}
\label{alg:hybridflow}
\begin{algorithmic}[1]
\Require Clients $C$, round $t$, $\lambda$, $\alpha_{\text{base}}$,
         $M_{\text{sync}}^{\min}$
\Statex \textbf{Phase 1: Estimation and Partitioning}
\State $\hat{\mathbf{T}} \gets \text{QuerySDN}(t,\, C)$
       \Comment{Estimate Request}
\ForAll{$c \in C$}
    \State $T_c \gets \hat{T}_{c,\text{S2C}} + \hat{T}_{c,\text{C2S}} +
           \bar{t}_c^{\text{comp}}$
\EndFor
\State $\mathcal{S}, \mathcal{A} \gets
    \text{HybridPartition}(C, \{T_c\}, \{w_c^{\text{imp}}\}, \lambda,
    M_{\text{sync}}^{\min})$
\Statex \textbf{Phase 2: Training}
\State Dispatch $\mathcal{A}$ to background training
\State $R_{\mathcal{S}} \gets \text{TrainSync}(\mathcal{S})$
\Statex \textbf{Phase 3: Two-Stage Aggregation}
\State $w_{\text{sync}} \gets \text{Aggregate}(R_{\mathcal{S}})$
       \Comment{Eq.~via configured strategy}
\State $w_{\text{global}}^{(t)} \gets
    \text{IntegrateAsync}(w_{\text{sync}},\, \text{Buffer})$
    \Comment{Eqs.~\ref{eq:alpha_eff}--\ref{eq:async_merge}}
\Statex \textbf{Phase 4: Update and Report}
\State Update $\{w_c^{\text{imp}}\}$ from $R$ and $w_{\text{global}}^{(t)}$
\State $\text{ReportMeasurements}(t,\, C)$
       \Comment{Measurement Report}
\State \Return $w_{\text{global}}^{(t)}$
\end{algorithmic}
\end{algorithm}

A key property of Algorithm~\ref{alg:hybridflow} is that both path assignment and partitioning
complete \emph{before} training begins, eliminating the measurement instability associated with
mid-round switching. The round execution unfolds in four phases.

\textbf{Phase~1: Estimation and Partitioning.}
The FL server issues an estimate request to the SDN controller, which runs the
\emph{Flow Scheduler} and returns calibrated per-client communication time estimates. Using
these estimates, each client's total completion time $T_c$ is computed per
Equation~\eqref{eq:Tc}. The \emph{Hybrid Partitioner} then minimizes the cost function
$J(\mathcal{S})$ (Equation~\ref{eq:J}) via a greedy sweep over clients sorted by $T_c$,
producing the optimal sync set $\mathcal{S}$ and async set $\mathcal{A}$.

\textbf{Phase~2: Training.}
The server dispatches all clients in $\mathcal{A}$ concurrently to background training and
blocks on the completion of all clients in $\mathcal{S}$. Asynchronous updates complete at
varying times and are deposited into the thread-safe async buffer as they arrive.

\textbf{Phase~3: Two-Stage Aggregation.}
Stage~1 aggregates the synchronous results $R_{\mathcal{S}}$ to produce the intermediate model
$w_{\text{sync}}$. Stage~2 computes a staleness-weighted average of all buffered async updates
per Equation~\eqref{eq:wavg_async} and merges it into $w_{\text{sync}}$ with the collective
coefficient of Equation~\eqref{eq:alpha_coll}, yielding the updated global model
$w_{\text{global}}^{(t)}$. Clients that do not complete within the current round remain
in-flight; their expected residual time naturally biases them toward asynchronous classification
in the next round's partitioning.

\textbf{Phase~4: Update and Report.}
The FL server updates gradient importance weights $\{w_c^{\text{imp}}\}$ from the parameter
differences observed across all contributing clients relative to $w_{\text{global}}^{(t)}$.
It then reports real per-client communication times to the SDN controller, which uses them
to update the EWMA correction factors per Equation~\eqref{eq:cf}, closing the feedback loop
and improving path-score calibration in subsequent rounds.

\section{Experimental Results}\label{sec:results}
The preceding sections established how \name leverages SDN-provided communication time estimates
to partition clients into synchronous and asynchronous groups before each round, balancing round
latency against staleness risk. The experiments in this section evaluate the resulting
orchestration system along four axes: (1) per-round efficiency and synchronization blocking
reduction, as the primary management outcome; (2) adaptivity of the sync/async partition to
per-round network conditions; (3) accuracy of the SDN-layer closed-loop predictor that drives
partitioning decisions; and (4) convergence speed, as evidence that the orchestration actions
preserve ML utility. Together, these experiments test whether the SDN-FL design choices
described in Section~\ref{sec:fl_layer} translate into measurable management and training
improvements under realistic network heterogeneity.

\begin{figure*}[t]
    \centering
    \begin{subfigure}[t]{0.48\textwidth}
        \includegraphics[width=\linewidth]{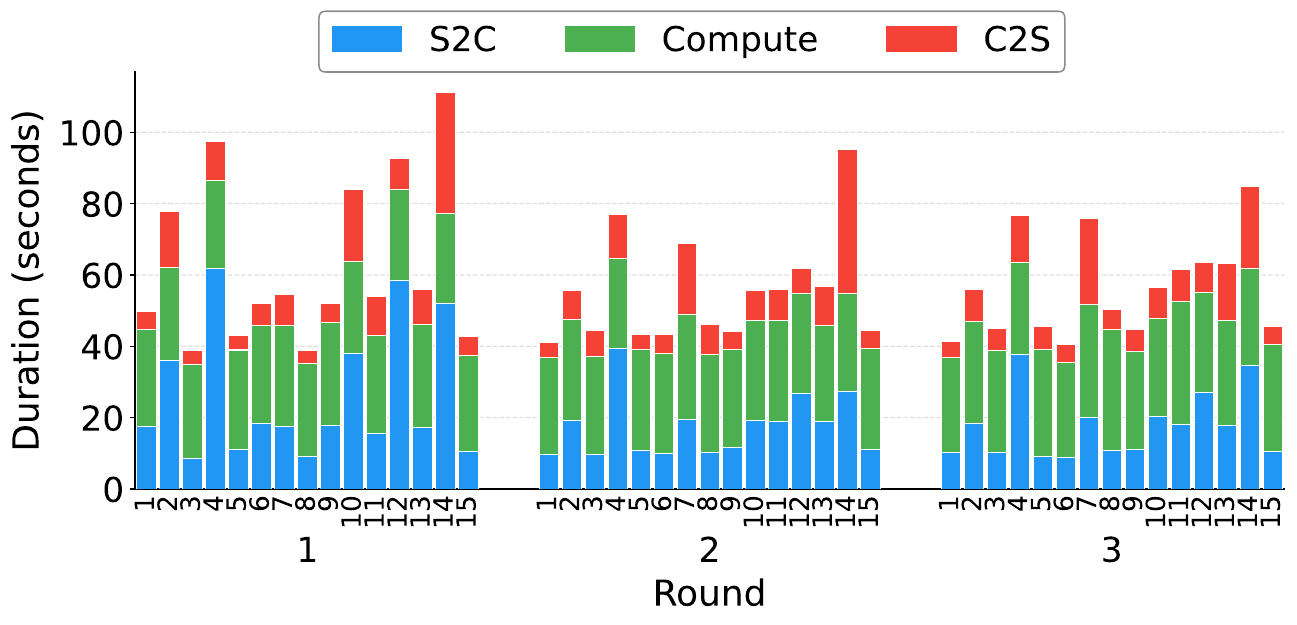}
        \caption{\smartflow}
    \end{subfigure}
    \hfill
    \begin{subfigure}[t]{0.48\textwidth}
        \includegraphics[width=\linewidth]{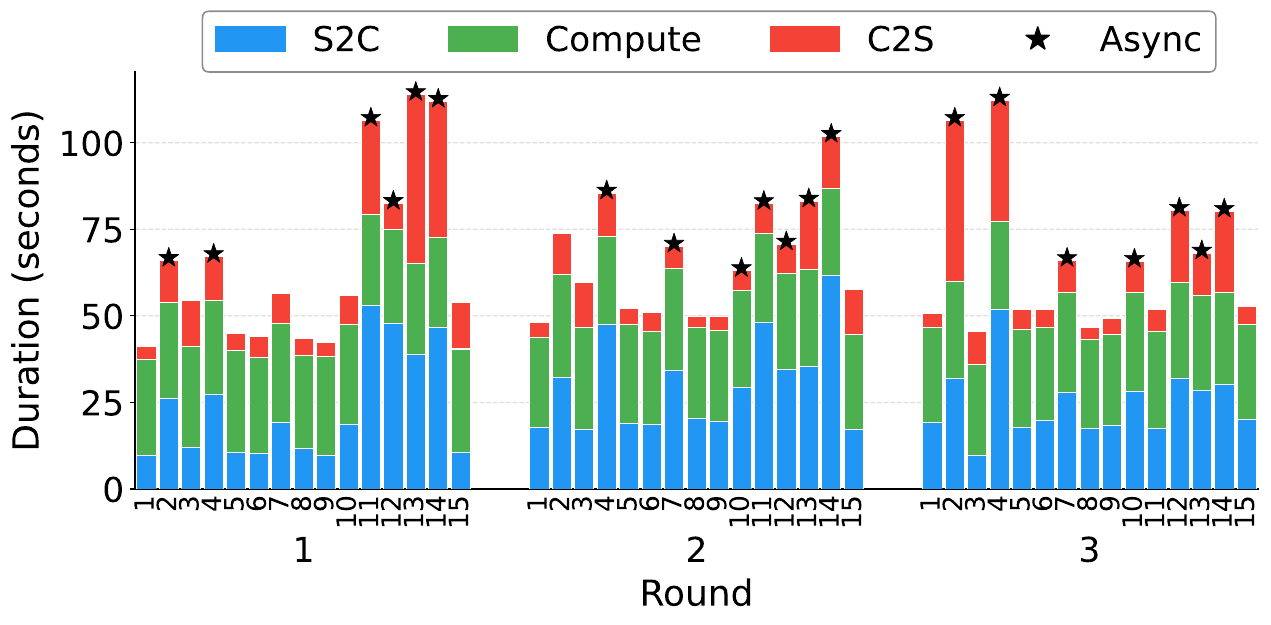}
        \caption{\name}
    \end{subfigure}
    \caption{Per-round stacked time breakdown for topology E1. Each bar decomposes one client's round into S2C transfer, local training, and C2S transfer; (a) shows \smartflow, (b) shows \name. Stars in (b) mark clients reclassified as asynchronous.}
    \label{fig:stacked_comm_bars}
\end{figure*}

\subsection{Experimental Setup}
\label{sec:setup}

All experiments use FLEET~\cite{testbed-paper} as the emulation framework and draw network topologies from the Gabriel collection in TopologyHub~\cite{gabriel}.
Three topologies of increasing scale define the evaluation scenarios: E1 with 15 switches and 15
FL clients, E2 with 25 switches and 25 clients, and E3 with 35 switches and 35 clients. All
links operate at 100~Mbps, and background traffic follows a Poisson arrival process that injects
time-varying congestion across the network. Every client runs on identical compute hardware, so
that observed performance differences arise exclusively from network heterogeneity. The hardware platform is a single CPU-only server equipped with 128~GB RAM and an AMD EPYC 9124
16-core, 32-thread processor.

All experiments use a fixed dataset and model to isolate the impact of network conditions on federated learning performance. CIFAR-10 is partitioned using \texttt{PathologicalPartitioner}, which assigns 7 of the 10 label classes to each client, creating a non-IID setting in which asynchronous FL is particularly susceptible to gradient staleness and contribution imbalance. MobileNetV3-Large~\cite{mobilenetlarge} is selected to generate realistic communication workloads; its 14.2~MB model size makes wide-area communication a significant contributor to round completion time, allowing the evaluation to capture the effects of network heterogeneity. As topology size increases, the training dataset is distributed across a larger number of clients (approximately 3,000 samples per client in E1, 2,000 in E2, and 1,500 in E3), while a fixed 10,000-sample server-side evaluation set is used across all experiments.

Emulation allows the same topology, traffic, and compute conditions to be replayed across
baselines, isolating the effect of network-aware partitioning while preserving realistic
control-plane interaction through ONOS and OpenFlow. All three baselines run the same
implementation stack (ONOS, Flower, ZeroMQ) and share identical hardware and background
traffic conditions, so observed differences arise exclusively from the partitioning strategy. 
% Additionally, since all client heterogeneity is network-induced by design, compute-aware hybrid baselines are not applicable; the relevant comparison is between approaches that differ in their use of network-layer information.

\begin{figure}[htbp]
    \centering
    \includegraphics[width=0.7\linewidth]{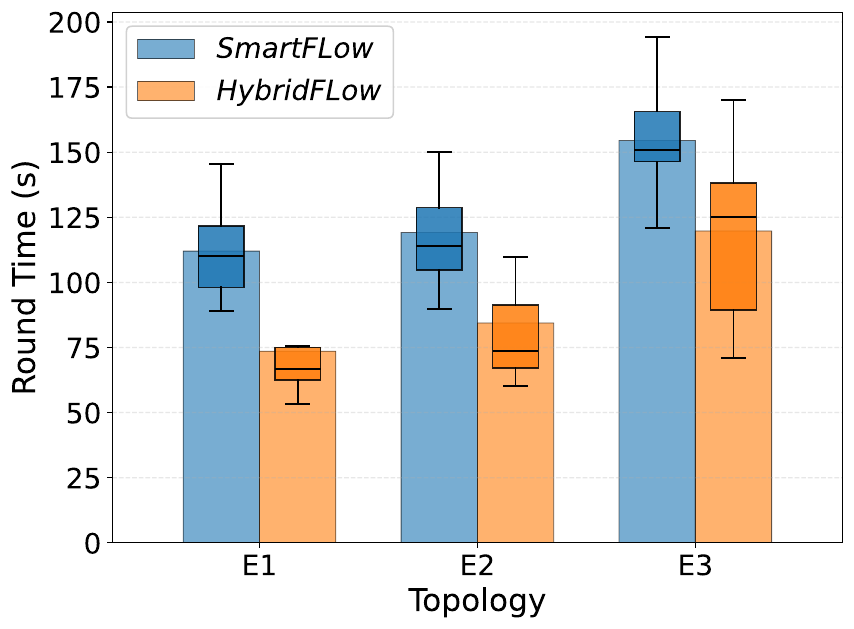}
    \caption{Average round time for \smartflow and \name across E1, E2, and E3. \name reduces synchronization blocking by removing network-induced tail clients from the synchronous phase, cutting average round time by 30-40~seconds in every topology.}
    \label{fig:avg_round_time}
    \vspace{-3mm}
\end{figure}

% Box-plot overlays show the median, interquartile range, and whiskers.

\begin{figure}[ht]
    \centering
    \includegraphics[width=0.7\linewidth]{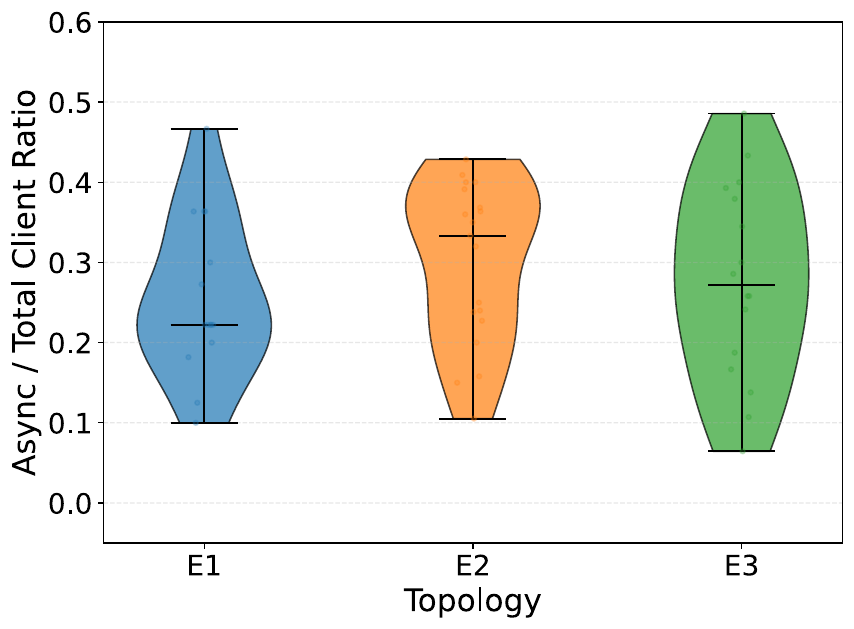}
    \caption{Distribution of per-round asynchronous-to-total client ratio for \name across E1, E2, and E3. Each violin encodes the full empirical distribution; a horizontal bar marks the median.}
    \label{fig:async_ratio_violin}
    \vspace{-3mm}
\end{figure}

Three training configurations are compared. \smartflow~uses SDN-optimized routing with fully synchronous training, while \fedasync~follows the FedAsync protocol~\cite{fedasync}. \name~combines synchronous and asynchronous participation through network-aware client partitioning. For consistency, \name~adopts the FedAvgM aggregation strategy used by \smartflow ($\texttt{server\_momentum}=0.9$) and the exponential staleness function used by \fedasync (\texttt{staleness\_alpha}=$0.25$, \texttt{mixing\_alpha}=$0.75$). Finally, \name~uses $\lambda = 0.35$ in Equation~(\ref{eq:J}), selected from preliminary experiments on E1. Across $\lambda \in \{0.1, 0.3, 0.5, 0.8\}$, performance varied smoothly, with $\lambda = 0.35$ providing a balanced trade-off between round latency and update staleness.

Training terminates when server evaluation accuracy reaches 80\%, making
\emph{time-to-target-accuracy} the primary metric. This criterion directly reflects how quickly
each method produces a usable model under identical network conditions. Any method that fails to
reach the target within the experiment budget is reported at its final achieved accuracy.

\begin{figure*}[htbp]
    \centering
    \begin{subfigure}[t]{0.31\textwidth}
        \includegraphics[width=\linewidth]{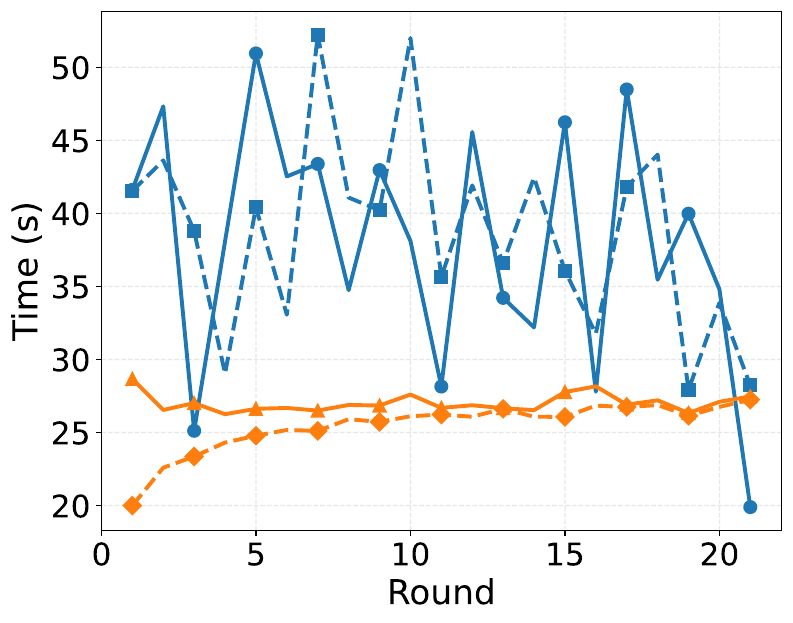}
        \caption{Topology E1}
    \end{subfigure}
    \hfill
    \begin{subfigure}[t]{0.31\textwidth}
        \includegraphics[width=\linewidth]{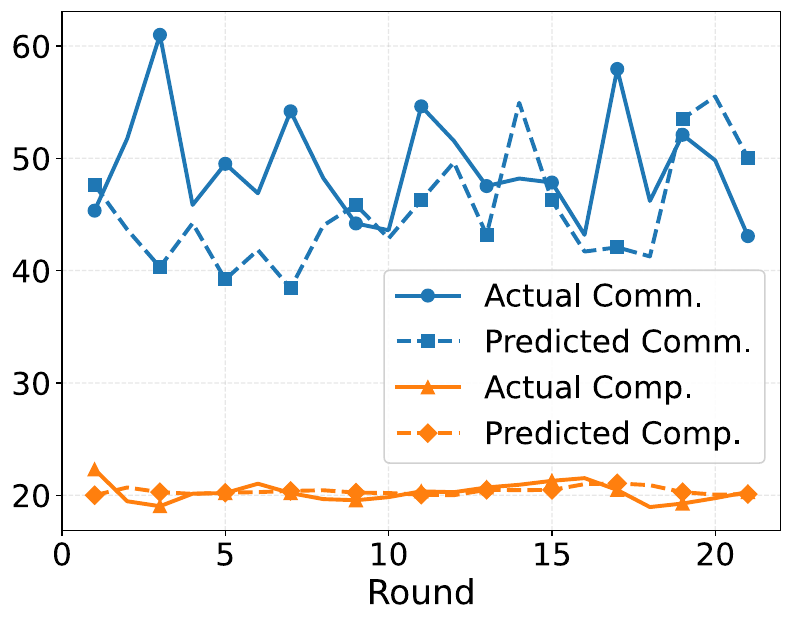}
        \caption{Topology E2}
    \end{subfigure}
    \hfill
    \begin{subfigure}[t]{0.31\textwidth}
        \includegraphics[width=\linewidth]{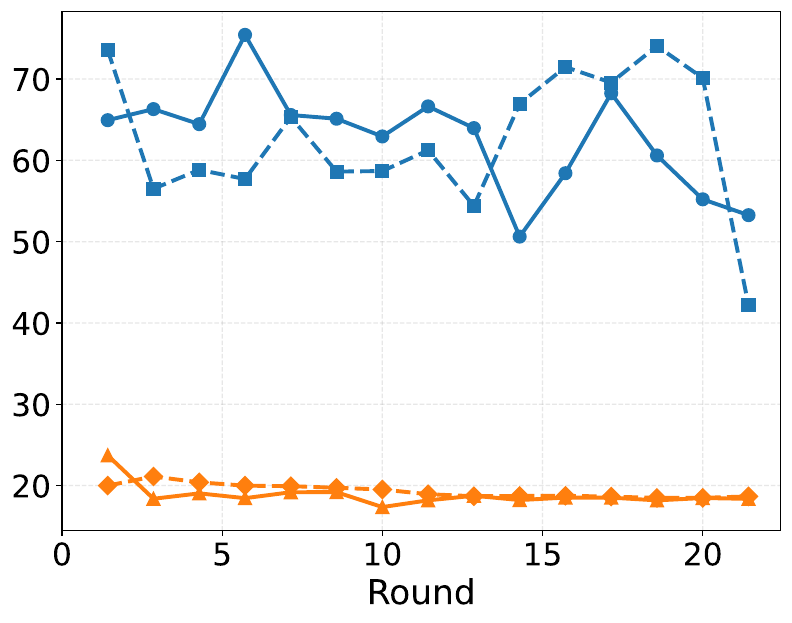}
        \caption{Topology E3}
    \end{subfigure}
    \caption{Predicted vs.\ actual communication and computing time for \name across E1,
    E2, and E3. Close tracking validates the closed-loop control mechanism: measurements
    reported after each round calibrate the SDN controller's future estimates, keeping
    prediction error low enough for reliable per-round orchestration.}
    \label{fig:pred_vs_actual}
    \vspace{-3mm}
\end{figure*}

\begin{figure*}[htbp]
    \centering
    \begin{subfigure}[b]{0.3\textwidth}
        \includegraphics[width=\linewidth]{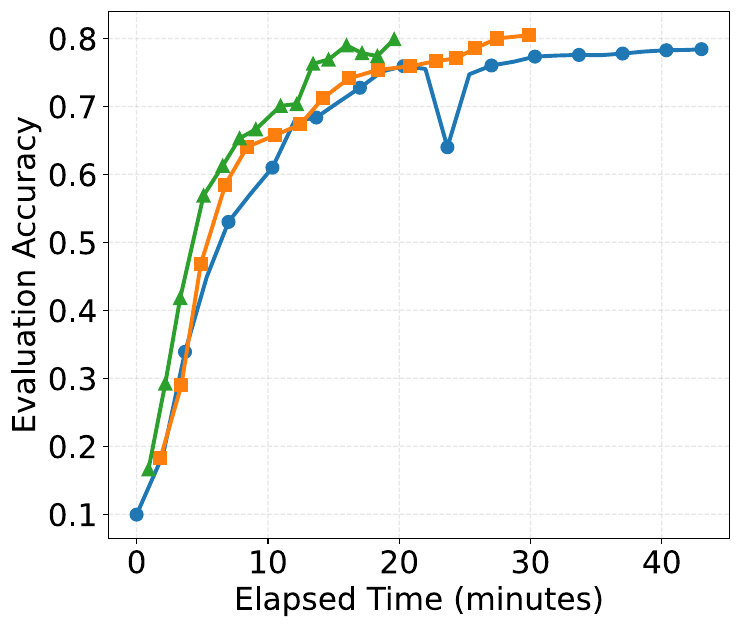}
        \caption{Topology E1}
    \end{subfigure}
    \begin{subfigure}[b]{0.3\textwidth}
        \includegraphics[width=\linewidth]{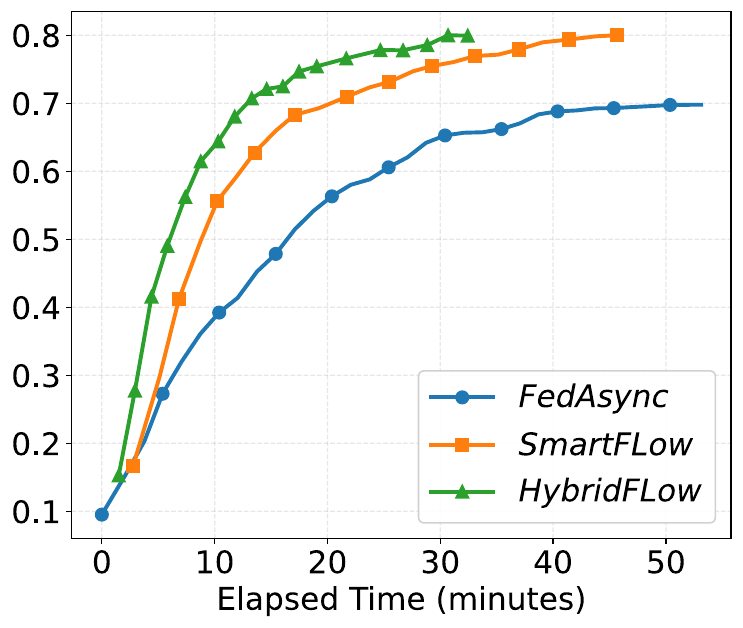}
        \caption{Topology E2}
    \end{subfigure}
    \begin{subfigure}[b]{0.3\textwidth}
        \includegraphics[width=\linewidth]{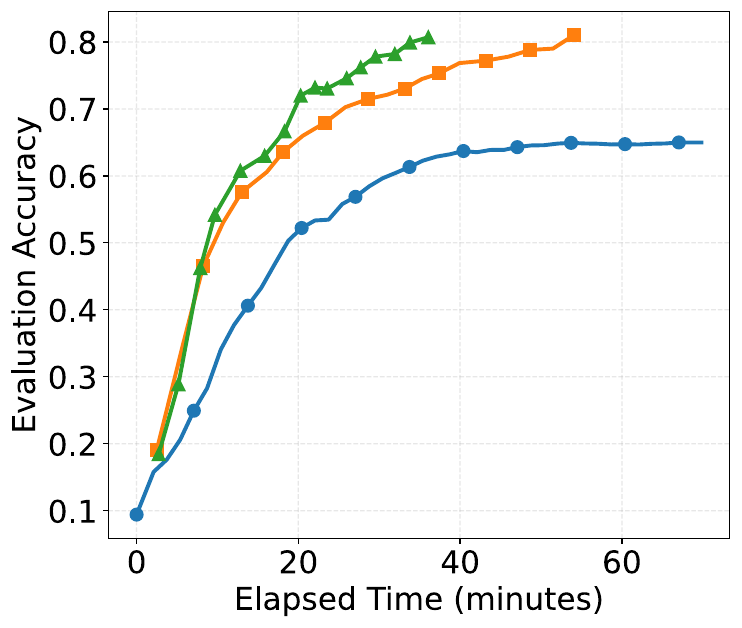}
        \caption{Topology E3}
    \end{subfigure}
    \caption{Server evaluation accuracy versus wall-clock time for topologies E1, E2, and E3. Each subplot compares \smartflow, \fedasync, and \name. \name reaches the 80\% target first in all three topologies, while \fedasync fails to reach the target in any topology.}
    \label{fig:accuracy_vs_time}
    \vspace{-3mm}
\end{figure*}

\subsection{Results} \label{sec:results_narrative}
The central outcome of \name~is the reduction of synchronization blocking during federated learning rounds. \figurename~\ref{fig:stacked_comm_bars} shows the mechanism: network transfer time is substantial for straggler clients, whose bars are visibly elongated by S2C and C2S segments.
\smartflow must wait each round for these slowest clients, whereas \name reclassifies them as
asynchronous (starred bars) and closes the round without them. Their updates are not discarded;
they are incorporated asynchronously in subsequent rounds, subject to a staleness penalty,
preserving their contribution while removing their latency from the synchronous phase. Per-client communication cost remains comparable between the two methods, indicating that the improvement arises from reduced synchronization delay rather than reduced communication volume.

\figurename~\ref{fig:avg_round_time} confirms this at the aggregate level: average round time
drops by 30--40~seconds for \name relative to \smartflow in every topology, a reduction that
is consistent across E1, E2, and E3. The box-plot spread for \name is narrower in E1 and E2,
reflecting more predictable round durations when stragglers are excluded; in E3, richer path
diversity at 35-client scale introduces greater round-to-round variability that offsets the
variance reduction.

\figurename~\ref{fig:async_ratio_violin} shows that the SDN control plane does not apply a
static policy: no violin collapses to a point across any topology. Distribution shape tracks
network heterogeneity---skewed toward low async fractions in E1, shifted higher in E2 as
greater path diversity increases reclassification frequency, and approximately symmetric in E3
where 35-client topology variability balances low- and high-reclassification rounds. The
correlation between distribution shape and topology scale confirms that the partitioner
continuously adjusts the synchronous boundary in response to per-round network conditions.

The closed-loop control mechanism is validated by \figurename~\ref{fig:pred_vs_actual}, which
overlays predicted and actual communication and computing times across all rounds and topologies.
Communication MAE is 7.2~s (E1), 6.2~s (E2), and 9.2~s (E3); computing MAE stays below
1.5~s in all three. The close tracking validates the EWMA correction mechanism
(Equation~\ref{eq:cf}), which calibrates the SDN controller's estimates using real measurements
reported after each round. Reliable prediction is a prerequisite for the round-time gains:
misclassifying clients---retaining stragglers synchronously or ejecting fast ones
prematurely---would negate the benefit.

The management actions preserve convergence quality. \figurename~\ref{fig:accuracy_vs_time}
validates this across all three topologies: \name reaches the 80\% accuracy target 33--40\%
faster than \smartflow in every configuration, a consistent advantage from the 15-client E1
topology through the 35-client E3 topology. \fedasync fails to reach the 80\% target in any
topology under non-IID data, confirming that unconstrained asynchronous aggregation accumulates stale
gradients that prevent convergence when data distributions are heterogeneous.

Taken together, the SDN-FL orchestration pipeline is coherent: accurate closed-loop time
prediction (\figurename~\ref{fig:pred_vs_actual}) enables reliable per-round partitioning
(\figurename~\ref{fig:async_ratio_violin}), which reduces synchronization blocking
(\figurename~\ref{fig:avg_round_time}) without sacrificing convergence quality
(\figurename~\ref{fig:accuracy_vs_time}). The net effect is a 33--40\% reduction in
time-to-target-accuracy compared to \smartflow, achieved without the convergence degradation
that affects \fedasync under non-IID data distributions.

\subsection{Limitations of the Study}
The evaluation is conducted in an emulated cross-silo environment rather than a production WAN deployment. While this enables analysis of network-aware partitioning with realistic SDN control-plane interactions, validation at larger scales and across operational WANs remains future work. \name assumes the availability of topology-wide SDN telemetry, which is appropriate for enterprise, campus, research or private networks but may not apply where network visibility is limited or routing decisions are controlled by external providers. Finally, the current design assumes trustworthy telemetry and well-behaved participants; the effects of telemetry inaccuracies, controller failures, and adversarial clients remain open directions for future work.

\section{Conclusion}
\name is a closed-loop SDN orchestration framework for Hybrid Federated Learning in cross-silo environments. Unlike existing hybrid FL approaches that rely primarily on compute-layer signals, \name incorporates topology-wide network telemetry into per-round sync/async partitioning decisions through an SDN control plane. By combining communication-time prediction, network-aware partitioning, and feedback-driven calibration within a unified orchestration loop, \name enables training decisions that account for both communication latency and update staleness. Experiments across emulated cross-silo topologies demonstrate that network-aware orchestration reduces synchronization blocking while preserving convergence quality under non-IID data distributions. Relative to \smartflow, \name reaches the 80\% accuracy target 33--40\% faster and reduces average round duration by 30--40~seconds across all evaluated topologies. These results demonstrate that topology-aware network telemetry can serve as an effective control signal for hybrid federated learning orchestration, enabling communication-aware training decisions in heterogeneous wide-area environments.

\section*{Acknowledgment}
The work in this study was supported in part by NSF grant 2427408 and 2451376. The authors used LLM tools (ChatGPT and Claude) for text formatting. All scientific content including experimental design, analysis, results, figures, and citations is the authors' own.

% \clearpage
\bibliographystyle{IEEEtran}
% \footnotesize
\bibliography{ref}

\end{document}